\documentclass[twocolumn,showpacs,preprintnumbers,amsmath,amssymb]{revtex4}

\usepackage{graphicx}
\usepackage{dcolumn}
\usepackage{bm}
\usepackage{epsfig}

\begin{document}

\preprint{FIS-UI-TH-04-01}

\title{Neutrino Electromagnetic Form Factor and Oscillation Effects on Neutrino 
Interaction With Dense Matter}

\author{C. K. Williams, P. T. P. Hutauruk, A. Sulaksono, T. Mart}
 \affiliation{Departemen Fisika, FMIPA, Universitas Indonesia, Depok 16424, Indonesia}

\date{\today}

\begin{abstract}
The mean free path of neutrino - free electron gas interaction has been 
calculated by taking into account the neutrino electromagnetic form 
factors and the possibility of neutrino oscillation. It is shown 
that the form factor effect becomes significant for a neutrino magnetic 
moment $\mu_{\nu}\geq10^{-10} \mu_B$ and for a neutrino radius  
$R\geq10^{-6}$ MeV$^{-1}$. The mean free path is found to be sensitive 
to the $\nu_e-\nu_\mu$ and  $\nu_e-\nu_e^R$  transition probabilities.

\end{abstract}



\maketitle

Neutrino interaction with dense matter plays an important role in astrophysics,
e.g., in the formation of supernova and the cooling of young neutron stars 
\cite{sanjay,Horo1,Horo66,niembro,mornas,leinson,fabbri,Horo2}. Earlier calculation on neutrino interactions with electrons gas,
dense and hot matter, based on the standard model has been performed by Horowitz and 
Wehrberger  \cite{Horo1,Horo66}. Some relativistic calculations of neutrino mean free path in hot and dense matter have been also done in Refs. \cite{niembro,mornas,leinson,fabbri}. Recently, due to a demand on a more realistic neutrino mean free path for supernova simulations, a mean free path calculation by taking into account the weak magnetism of nucleons has been also performed \cite{Horo2}. 

However, certain phenomena such as solar neutrinos, atmospheric neutrinos problems, 
and some astrophysics and cosmology arguments need explanations beyond the 
standard model assumption of neutrino's properties such as neutrino 
oscillation \cite{kuo,pulido}, the helicity flipping of neutrinos \cite{ayala,enqvist,gaemers,grimus1} and neutrino electromagnetic form factors. We note that the upper bound of the neutrino magnetic moment extracted from the Super-Kamiokande solar data \cite{liu,beacom} falls in the range of $(1.1-1.5)\times 10^{-10} \mu_B$, where $\mu_B=e/2m_e$ stands for the Bohr magneton. Other experimental limits \cite{grimus,dara} give $\mu_{\nu}$ $<$ 1.0 $\times 10^{-10}\mu_B$, whereas signals from Supernova 1987A (SN1987A) require that $\mu_{\nu}$ $\le$ 1.0 $\times 10^{-12}\mu_B$. These bounds have been derived by considering the helicity flipping neutrino scattering in a supernova core \cite{nuno}. In the case of random magnetic fields inside the sun, one can obtain a direct constraint on the neutrino magnetic moment of  $\mu_{\nu}$ $\le$ 1.0 $\times 10^{-12}\mu_B$, similar to the bounds obtained from the star cooling \cite{mira}. In addition, data from muon neutrino- and anti neutrino-electron scatterings  \cite{allen,Kerimov} and a close examination to the data over the years from Kamiokande II and Homestake according to Mour$\tilde{a}$o {\it et al.} \cite{Mourao}, similarly give a neutrino average squared radius $R^2 \sim$ 25 $\times 10^{-12}$ MeV$^{-2}$ with $R^2$= $\langle{R_V^2}\rangle$ + $\langle{R_A^2}\rangle$. The definitions of $\langle{R_V^2}\rangle$ and $\langle{R_A^2}\rangle$ will be explained later.

Therefore, in connection with the demand on realistic neutrino mean free path in dense and hot matter, an extension of the previous study \cite{niembro,mornas,leinson,fabbri, Horo2} which takes 
into account the electromagnetic form factors of neutrinos and neutrino oscillations is inevitable. As a first step before that, in this report we calculate the mean free path of neutrino-free electrons gas where those effects are included. Here we assume that neutrinos are massless and the RPA correlations can be neglected. Furthermore, we use zero temperature approximation in this calculation.

In the standard model, where the momentum transfer is much less than the $W$ 
mass, direct $Z^0$ and $W^{\pm}$ contributions to the matrix element 
$\mathcal{M}$ can be written as an effective four-point coupling 
\cite{Horo66,Kerimov}
\begin{equation}
\mathcal{M}_W=\frac{G_F}{\sqrt2}[\bar{U}(k')\gamma^{\mu}(1+\gamma^5)U(k)]
[\bar{U}(p')J_{\mu}U(p)],
\end{equation}
where $G_F$ is the coupling constant of weak interaction, 
$U(k)$ and $U(p)$ are neutrino and electron spinors, respectively, 
and the current $J_\mu$ is defined by
\begin{equation}
J_\mu=\gamma_\mu(C_V+C_A\gamma^5).
\end{equation}
The vector and axial vector couplings $C_V$ and $C_A$ can be written in 
terms of Weinberg angle $\theta_{W}$ (where $\sin^2\theta_{W}\approx 0.223$ 
\cite{Horo66,niembro}) as $C_V=2\sin^2\theta_{W}\pm1/2$ and $C_A=\pm 1/2$ 
(the upper sign is for $\nu_e$, the lower sign is for $\nu_{\mu}$ and 
$\nu_{\tau}$).

The electromagnetic properties of Dirac neutrinos are described in 
terms of four form factors, i.e., $f_{1\nu}, g_{1\nu}, f_{2\nu}$ and 
$g_{2\nu}$, which stand for the Dirac, anapole, magnetic, and electric 
form factors, respectively. The matrix element for the neutrino-electron 
interaction which contains electromagnetic form factors reads \cite{Kerimov}
\begin{eqnarray}
\mathcal{M}_{EM} &=& \frac{4\pi\alpha}{q^2}[\bar{U}(p')\gamma_{\mu}U(p)]
\Biggl\{ \bar{U}(k')\Biggl[f_{m\nu}\gamma^{\mu}\nonumber\\
&+& g_{1\nu}\gamma^{\mu}\gamma^{5}-(f_{2\nu}+ig_{2\nu}\gamma^5)
\frac{P^\mu}{2m_e}\Biggr] U(k) \Biggr\}.
\end{eqnarray}
where $f_{m\nu}=f_{1\nu}+(m_\nu / m_e)f_{2\nu}$, 
$P^{\mu} = k^{\mu} +k^{\mu\prime}$, $m_\nu$ and $m_e$ are neutrino and 
electron masses, respectively.
In the static limit, the reduced Dirac form factor $f_{1\nu}$ and the 
neutrino anapole form factor $g_{1\nu}$ are related to the vector and 
axial vector charge radii $\langle{R^2_V}\rangle$ and $\langle{R^2_A}\rangle$ 
through \cite{Kerimov} 
\begin{equation}
f_{1\nu}(q^2)=\frac{1}{6} \langle{R^2_V}\rangle q^2 \quad\textrm{and}\quad 
g_{1\nu}(q^2)=\frac{1}{6}\langle{R^2_A}\rangle q^2.
\end{equation}
In the limit of $q^2\to0$, $f_{2\nu}$ and $g_{2\nu}$ define respectively 
the neutrino magnetic moment $\mu_{\nu}^{m}=f_{2\nu}(0)\mu_B$ and the 
(CP violating) electric dipole moment $\mu_{\nu}^e=g_{2\nu}(0)\mu_B$
 \cite{nardi,Kerimov}. Here we use  $\mu_{\nu}^2$=$\mu_{\nu}^{m ~ 2}$+ $\mu_{\nu}^{e ~ 2}$.

Next, we can obtain the differential cross section per volume $V$ 
for scattering of neutrinos with the initial energy $E_{\nu}$ and 
final energy $E_{\nu}'$ on the electrons gas. It consists of the contributions 
from weak (W) interaction, electromagnetic (EM) interaction, as well as
their interference (INT) term, i.e., 
\begin{widetext}
\begin{eqnarray}
\Biggl(\frac{1}{V}\frac{d^3\sigma}{d^2\Omega'dE_\nu'}\Biggr)_{\nu_e}= 
-\frac{1}{16\pi^2}\frac{E_\nu'}{E_\nu}\left[ \Biggl( \frac{G_F}{\sqrt{2}} 
\right)^2L^{\mu\nu}_\nu\Pi^{\rm Im (W)}_{\mu\nu}
+ \left( \frac{4\pi\alpha}{q^2}\right) ^2L^{\mu\nu}_\nu\Pi^{\rm Im (EM)}_{\mu\nu}+ 
\frac{8G_F\pi\alpha}{q^2\sqrt{2}}L^{\mu\nu}_{\nu}\Pi^{\rm Im (INT)}_{\mu\nu} \Biggr].
\label{1}
\end{eqnarray}
\end{widetext}
For each contribution, the neutrino tensors are given by 
\begin{widetext}
\begin{eqnarray}
L^{\mu\nu {\rm (W)}}_{\nu}  = 8[2k^{\mu}k^{\nu}-(k^{\mu}q^{\nu}+k^{\nu}q^{\mu})+g^{\mu\nu}(k \cdot q)
&-&i\epsilon^{\alpha\mu\beta\nu}k_{\alpha}k_{\beta}'],
\end{eqnarray}
\begin{eqnarray}
L^{\mu\nu {\rm (EM)}}_{\nu} &=& 4(f^{2}_{m\nu}+g^{2}_{1\nu})[2k^{\mu}k^{\nu}-(k^{\mu}q^{\nu}+k^{\nu}q^{\mu})
+ g^{\mu\nu} (k \cdot q)]-8if_{m\nu}g_{1\nu}\epsilon^{\alpha\mu\beta\nu}(k_{\alpha}k_{\beta}')\nonumber\\
&-&\frac{f^{2}_{2\nu}+g^{2}_{2\nu}}{m_e^2}(k \cdot q) 
 [4k^{\mu}k^{\nu}-2 (k^{\mu}q^{\nu}+q^{\mu}k^{\nu})+q^{\mu}q^{\nu}],
\end{eqnarray}
\begin{eqnarray}
L^{\mu\nu {\rm (INT)}}_{\nu} = 4(f_{m\nu}+g_{1\nu})[2k^{\mu}k^{\nu}-(k^{\mu}q^{\nu}+k^{\nu}q^{\mu})
+ g^{\mu\nu}(k \cdot q)-i\epsilon^{\alpha\mu\beta\nu}k_{\alpha}k_{\beta}'] ,
\end{eqnarray}
\end{widetext}
whereas the polarizations read
\begin{eqnarray}
\Pi^{\rm Im (W)}_{\mu\nu}&=&C^2_V\Pi^{{\rm Im}V}_{\mu\nu}+2C_VC_A\Pi^{{\rm Im}(V-A)}_{\mu\nu}
+C^2_A\Pi^{{\rm Im}A}_{\mu\nu},\nonumber\\
\Pi^{\rm Im (EM)}_{\mu\nu}&=&\Pi^{{\rm Im}V}_{\mu\nu},\nonumber\\
\Pi^{\rm Im (INT)}_{\mu\nu}&=&C_V\Pi^{{\rm Im}V}_{\mu\nu}+C_A\Pi^{{\rm Im}(V-A)}_{\mu\nu}.
\end{eqnarray}

Due to the current conservation and translational invariance, the vector 
polarization $\Pi^{{\rm Im}{V}}_{\mu\nu}$ consists of two independent 
components which we choose to be in the frame of 
$q^\mu \equiv(q_0,\vert{\vec{q}}\vert,0,0)$, i.e., 
\begin{eqnarray}
\Pi^{\rm Im V}_{T}&=&\Pi^V_{22}=\Pi^V_{33} \quad\textrm{and}\nonumber\\
\Pi^{\rm Im V}_{L}&=&-(q^2_{\mu}/\vert\vec{q}\vert^2)\Pi^V_{00}.\nonumber 
\end{eqnarray}
The axial-vector and the mixed pieces are found to be
\begin{eqnarray}
\Pi^{\rm Im (V-A)}_{\mu\nu}(q)=i\epsilon_{\alpha\mu 0\nu}q_{\alpha}\Pi_{VA},
\end{eqnarray}
and
\begin{eqnarray}
\Pi^{{\rm Im} A}_{\mu\nu}(q)=\Pi^{{\rm Im}V}_{\mu\nu}(q)+g_{\mu\nu}\Pi_A.
\end{eqnarray}
The explicit forms of $\Pi^V_{22}$, $\Pi^V_{00}$, $\Pi_{VA}$ and $\Pi_A$ 
are given in Ref.~ \cite{Horo66}. 
Thus the  analytical form of Eq.~(\ref{1}) can be obtained from the 
contraction of every polarization and neutrino tensors couple ($L^{\mu\nu }\Pi_{\mu\nu}$). 

If we take into account the possibility of the $ \nu_e-\nu_\mu$ transition, 
the cross section can be written in the form of \cite{HP,Morgan}

\begin{eqnarray}
\frac{d^3\sigma}{d^2\Omega'dE'}=P_{ee}\Biggr(\frac{d^3\sigma}{d^2\Omega'dE'}
\Biggl)_{\nu_e}+(1-P_{ee})\Biggr(\frac{d^3\sigma}{d^2\Omega'dE'}\Biggl)_{\nu_\mu}.\nonumber\\
\label{2}
\end{eqnarray}

Here $({d^3\sigma}/{d^2\Omega'dE'})_{\nu_e}$ is the cross section of the 
$\nu_e-e$ scattering. If $C_V$ and $C_A$ are replaced with $C_V-1$ and $C_A-1$, 
respectively, then the cross section becomes 
$({d^3\sigma}/{d^2\Omega'dE'})_{\nu_\mu}$, i.e., the cross section of  the 
$\nu_\mu-e$ scattering. $P_{ee}$ is the $\nu_e$'s flavor survival probability 
as a function of the neutrino energy.

Due to the assumption of massless neutrino, the $\nu_e$ helicity flip from 
left- to right-handed is only possible through it's dipole moment. Thus, 
the cross section after taking into account this possibility 
($\nu_e-\nu_e^R$ transition) reads \cite{grimus}
\begin{equation}
\frac{d^3\sigma}{d^2\Omega'dE'}=(1-P_{LL})\Biggr(\frac{d^3\sigma}{d^2\Omega'dE'}\Biggl)_{LR}
+ P_{LL}\Biggr(\frac{d^3\sigma}{d^2\Omega'dE'}\Biggl)_{\nu_e}.
\label{3}
\end{equation}
where $({d^3\sigma}/{d^2\Omega'dE'})_{_{LR}}$ is the $\nu_e-e$ scattering 
via neutrino dipole moment and $P_{LL}$ is the probability of $\nu_e$  
to be still left handed. 

\begin{figure}[t]
\centering
{\epsfig{file=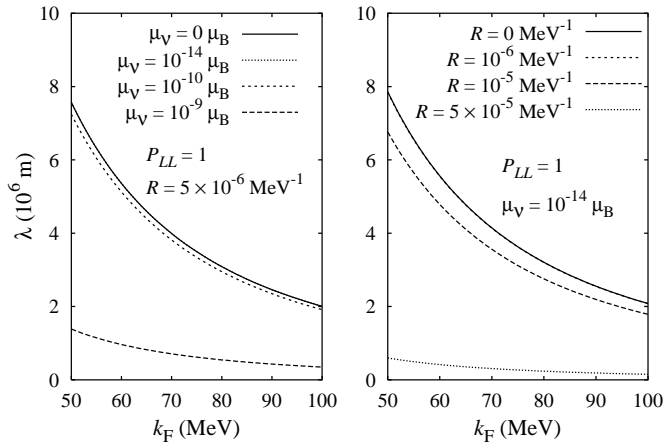,height=6.0cm}}
\caption{Total mean free path compared to the mean free path of weak interaction with various neutrino magnetic moments $\mu_{\nu}$ and radii $R$ as a function of Fermi momentum $k_F$. In the left panel the neutrino charge radius is fixed, while the neutrino magnetic moment is varied. In the right panel, we fix the neutrino magnetic moment, but vary the neutrino radius.}\label{fig1}
\end{figure}

\begin{figure}
\centering
{\epsfig{file=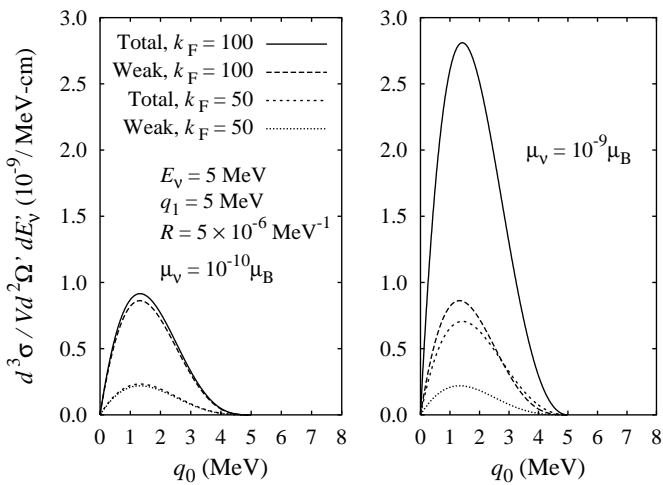,height=6.5cm}}
\caption{Total cross section compared to the cross section of weak interaction  as a function of energy transfer $q_0$ where momentum transfer $q_1$ is fixed. Here,  two different neutrino magnetic moments $\mu_{\nu}$  and Fermi momenta with a same neutrino charge radius are used. In the left panel we use  $\mu_{\nu}= 10^{-10}\mu_B$, while in the right panel $\mu_{\nu}= 10^{-9}\mu_B$.}\label{fig1a}
\end{figure}

Finally we can compute the mean free path from Eqs. (\ref{1}), (\ref{2}),
and (\ref{3}), by using 
\begin{eqnarray}
\frac{1}{{\lambda}(E_{{\nu}})} = \int_{q_{0}}^{2E_{{\nu}}-q_{0}}d|{\vec{q}}|\int_{0}^{2E_{{\nu}}}dq_{0}\frac{|{\vec{q}}|}{E'_{{\nu}}E_{{\nu}}}
2{\pi}\frac{1}{V}\frac{d^3{\sigma}}{d^2{\Omega}'dE'_{{\nu}}}.\nonumber\\
\end{eqnarray} 
In this calculation we use a neutrino energy of 5 MeV.

Figure\,\ref{fig1} shows the total mean free path compared to the mean free 
path of weak interaction with various neutrino effective moments $\mu_{\nu}$, and neutrino charge radii $R$.  
The total mean free path is the coherent sum of the weak, 
electromagnetic and the interference contributions.

There are also evidences that $R^2\approx 10^{-32}$ cm$^{-2}$ 
 or $R^2 \approx 25\times 10^{-12}$ MeV$^{-2}$ \cite{allen,Kerimov,Mourao}. 
Therefore, in the left panel of Fig.\,\ref{fig1} we use 
$R=5\times10^{-6}$ MeV$^{-1}$ and  vary $\mu_{\nu}$ between 
0 and $10^{-9}\mu_B$. In the right panel, 
we use $\mu_{\nu}=10^{-12}\mu_B$ as the strongest bound on the neutrino magnetic moment while $R$ is varied between
0 and $5\times10^{-5}$ MeV$^{-1}$. 

It is evident from the left panel of Fig.\,\ref{fig1} that for fixed $R$, 
the mean free path increases rapidly 
only after $\mu_{\nu}= 10^{-10}\mu_B$. As we can see from  Fig.\,\ref{fig1a} this increment is due to the significant difference between total and weak cross sections starting from $\mu_{\nu}= 10^{-10}\mu_B$. The summation of the longitudinal and transversal terms of the electromagnetic contribution is responsible for this. The right panel shows that for fixed $\mu_{\nu}$, the 
total mean free path and the mean free path of weak interaction 
show significant variance for $R\geq10^{-6}$ MeV$^{-1}$. This is also due to the fact that the summation of the longitudinal and transversal terms of the electromagnetic part of the cross section increases rapidly starting at  $R=10^{-6}$ MeV$^{-1}$.

\begin{figure}[t]
\centering
{\epsfig{file=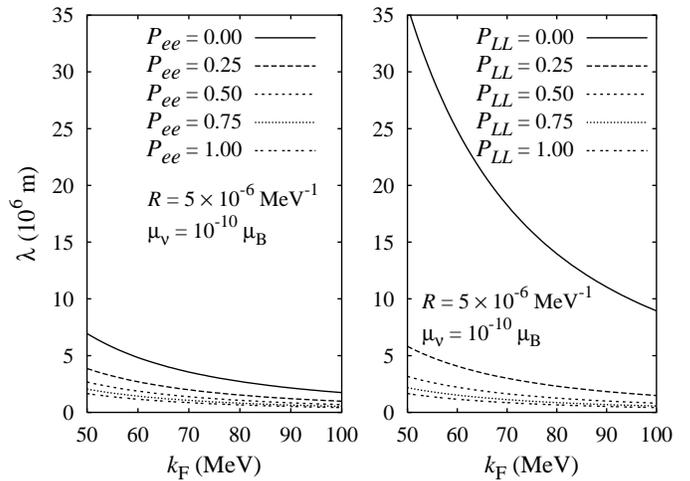,height=6.5cm}}
\caption{Total mean free path of $\nu_e$ that allows for $\nu_e-\nu_\mu$ and $\nu_e-\nu_e^R$ transitions with various $P_{LL}$ and $P_{ee}$ as a function of Fermi momentum $k_F$. In the left panel we vary the $\nu_e$'s flavor survival probability, while in the right panel the helicity flipping probability of neutrino is varied. }\label{fig2}
\end{figure}

Figure\,\ref{fig2} shows the effects of neutrino oscillations 
on the neutrino mean free path. In this case we do not calculate 
the  transition probabilities. Instead, we only study the variation of 
neutrino mean free path with respect to the transition probabilities 
of a left handed massless neutrino electron, $\nu_e$, oscillates 
to a left handed massless neutrino muon, $\nu_\mu$, or flips to 
a right handed neutrino electron, $\nu_e^R$. 

By comparing the possibility of $\nu_e-\nu_\mu$ transition (left panel of 
Fig.\,\ref{fig2}) and $\nu_e-\nu_e^R$ transition (right panel), we can 
clearly see that these effects lengthen the neutrino mean free path, 
where the rate depends on their survival probabilities. 
For smaller $P_{LL}$ (large flipping possibility), the path increment  
becomes more significant. This effect can be traced back to the 
value of $({d^3\sigma}/{d^2\Omega'dE'})_{_{LR}}$ in Eq.\,(\ref{3}) 
which is smaller 
than that of $({d^3\sigma}/{d^2\Omega'dE'})_{\nu_e}$. On the other hand, 
for small $P_{ee}$ the possibility of $\nu_e-\nu_\mu$ oscillation does not 
change the neutrino mean free path dramatically. This fact arises 
because the difference between $({d^3\sigma}/{d^2\Omega'dE'})_{\nu_\mu}$ 
and $({d^3\sigma}/{d^2\Omega'dE'})_{\nu_e}$ in Eq.\,(\ref{2}) is
not as large as in the case of Eq.\,(\ref{3}). Therefore different from the mean free path with flavor changing possibility, the mean free path with helicity flipping possibility depends strongly on the value of  
$\mu_{\nu}$. For example we have also found that with decreasing $P_{LL}$ the mean free path grows more rapidly when we use $\mu_{\nu}= 10^{-12}\mu_B$ rather than 
$\mu_{\nu}= 10^{-10}\mu_B$.  

In conclusion, we have studied the sensitivity of the neutrino mean free 
path to the neutrino electromagnetic form factors and neutrino oscillations. 
It is found that the electromagnetic form factor has a significant role 
if $\mu_{\nu}\geq10^{-10} \mu_B$ and $R\geq10^{-6}$  MeV$^{-1}$. We note that these values are larger than their largest upper bounds. 
It would be interesting to see whether or not such phenomenon would also appear
if contributions from the neutrino-nucleon scatterings were taken into account.
Future calculation should address this question.
The mean free path is also found to be sensitive to the neutrino oscillations and 
depends on the transition probabilities of $\nu_e-\nu_\mu$ and $\nu_e-\nu_e^R$.
This result clearly indicates that realistic mean free path calculations in the future
should be performed with appropriate values of the 
$\nu_e-\nu_\mu$ and $\nu_e-\nu_e^R$ transition probabilities.

TM and AS acknowledge the support from the QUE project.


\begin{thebibliography}{99}
\bibitem{sanjay}
     Sanjay Reddy, M. Prakash, J. M. Lattimer, J. A. Pons,
     Phys.\ Rev.\ C {\bf 59}, 2888 (1999).
\bibitem{Horo1}
      C. J. Horowitz and K. Wehrberger,
      Phys. Rev. Lett. {\bf 66}, 272 (1991);
\bibitem{Horo66}
     C. J. Horowitz and K. Wehrberger, Nucl. Phys. {\bf A531}, 665 (1991);  Phys. Lett. B {\bf 226}, 236 (1991). 
\bibitem{niembro}
    R. Niembro, P. Bernados, M. L$\acute{o}$pez-Quele, S. Marcos, Phys. Rev. C
    {\bf 64}, 055802 (2001).
\bibitem{mornas}
    L. Mornas and A. Parez, Eur. Phys. J. A {\bf 13}, 383 (2002).
\bibitem{leinson}
    L. B. Leinson, Nucl. Phys. A {\bf 707}, 543 (2002).
\bibitem{fabbri}
    G. Fabbri and F. Matera, Phys. Rev. C {\bf 54}, 2031 (1996).
\bibitem{Horo2}
    C. J. Horowitz and M. A. P$\acute{e}$rez-Garc$\acute{i}$a, Phys. Rev. C
    {\bf 68}, 025803 (2003).
\bibitem{kuo}
     T. K. Kuo and J. Pantaleone, Rev. Mod. Phys. {\bf 61}, 937 (1992); and references therein.
\bibitem{pulido}
     J. Pulido, Phys. Rept. {\bf 211}, 167 (1992); and references therein.
\bibitem{ayala}
     A. Ayala, J. C. D'Olivo and M. Tores, Phys. Rev. D {\bf 59}, 111901(1999).
\bibitem{enqvist}
     K. Enqvist, P. Keraenen and J. Maalampi, Phys. Lett. B {\bf 438}, 295 (1998).
\bibitem{gaemers}
     K. J. F. Gaemers, R. Gandhi, J. M. Lattimer, Phys. Rev. D {\bf 40}, 309 (1989).
\bibitem{grimus1}
    W. Grimus, P. Stockinger, Phys.\ Rev.\ D {\bf 57}, 1762 (1998).
\bibitem{liu}
     D. W. Liu {\it et al.}, Phys. Rev. Lett {\bf 93}, 021802 (2004).
\bibitem{beacom}
     J. F. Beacom and P. Vogel, Phys. Rev. Lett {\bf 83}, 5222 (1999).
\bibitem{grimus}
     W. Grimus {\it et al.}, Nucl. Phys. B {\bf 648}, 376 (2003).
\bibitem{dara}
     Z. Daraktchieva {\it et al.}, Phys. Lett B {\bf 564}, 190 (2003).
\bibitem{nuno}
     N. Nunokawa, R. Tom$\acute{a}$s and J. W. F. Valle, Astropart. Phys. {\bf 11}, 317 (1999); and references therein.
\bibitem{mira}
     O. G. Miranda, T. I. Rashba, A. I. Rez and J. W. F. Valle, Phys. Rev. Lett {\bf 93}, 051304 (2004).
\bibitem{allen}
     R. C. Allen {\it et al.}, Phys. Rev. D {\bf 43}, 1 (1991).
\bibitem{Kerimov}
    B. K. Kerimov, M. Ya Safin and H. Nazih, Izv.\ Ross.\ Akad.\ Nauk. SSSR. Fiz.  {\bf 52}, 136 (1998).
\bibitem{nardi}
     Enrico Nardi, AIP Conf.\ Proc.\  {\bf 670}, 118 (2003).
\bibitem{Mourao}
    A. M. Mour$\tilde{a}$o, J. Pulido, and J. P. Ralston, Phys. Lett. B {\bf 285}, 364 (1992).
\bibitem{HP}
    H. P. Simanjuntak and A. Sulaksono, Mod. Phys. Lett. {\bf 9A}, 2179 (1994);
    A. Sulaksono and H. P. Simanjuntak, Solar Phys. {\bf 151}, 205 (1994).
\bibitem{Morgan}
    J. Morgan, Phys. Lett. B{\bf 102}, 247 (1981); M. Fukugita and S. Yazaki, 
    Phys. Rev. D {\bf 36}, 3817 (1987).



\end{thebibliography}
\end{document}